%% file: gmm.tex
\documentclass{LMCS}

\usepackage{amsmath,hyperref}

\input{definitions.tex}

\input{defpart.tex}

\def\doi{2 (4:1) 2006}
\lmcsheading%
{\doi}
{1--15}
{}
{}
{Jan.~\phantom{0}4, 2006}
{Sep.~27, 2006}
{}   

\begin{document}

\title{Generalized Majority-Minority Operations are Tractable}

\author[V.~Dalmau]{V\'{\i}ctor Dalmau}
\address{Departament de Tecnologia, Universitat Pompeu Fabra, \\
Estaci\'o de Fran\c{c}a, Passeig de la Circumval.lacio 8. Barcelona 08003, Spain}
\email{victor.dalmau@tecn.upf.es}

\keywords{constraint satisfaction, complexity, Mal'tsev, near-unanimity}
\subjclass{F.4.1}

\begin{abstract}

  Generalized majority-minority (GMM) operations are introduced as a
  common generalization of near unanimity operations and Mal'tsev
  operations on finite sets.  We show that every instance of the
  constraint satisfaction problem (CSP), where all constraint
  relations are invariant under a (fixed) GMM operation, is solvable in
  polynomial time. This constitutes one of the largest tractable cases
  of the CSP.
\end{abstract}

\maketitle

\section{Introduction}

Constraint satisfaction problems arise in a wide variety of domains,
such as combinatorics, logic, algebra, and artificial intelligence.
An instance of the constraint satisfaction problem (CSP)
consists of a set of variables, a set of values (which can be taken
by the variables), called domain,  and a set of constraints,
where a constraint is a pair given by a list of variables, called
scope,  and a relation indicating the valid combinations of values for
the variables in the scope;
the goal is to decide whether or not there is an assignment of values to the
variables satisfying all of the constraints.
It is well known that the CSP admits several different but equivalent definitions.
Feder and Vardi~\cite{Feder/Vardi:1998} formulated it as the problem of deciding whether
there
exists an homomorphism between two given relational structures. Also, an instance
of the CSP can be viewed as a positive primitive sentence; the question
is to decide whether or not the sentence is true.

In its full generality the CSP is NP-complete. This fact motivates
the project of identifying restricted subclasses of the problem that
are solvable in polynomial time. The most customary way to restrict
the CSP is by fixing a set of relations $\Gamma$, generally called
{\em constraint language} or {\em basis} and consider only instances
of the CSP in which every relation appearing in a constraint belongs
to $\Gamma$; this restricted version of the problem is generally
denoted by $\csp(\Gamma)$. Much effort has been devoted to the goal
of isolating those constraint languages, $\Gamma$, for which its
associated constraint satisfaction problem, $\csp(\Gamma)$, is
polynomial-time solvable. Despite the large amount of results in
this
direction~\cite{Bulatov:3element,Bulatov:maltsev,Bulatov03:conservative,Bulatov04:graph,Bulatov:2semilattices,Bulatov/Jeavons/Volkov:2001,Bulatov/Krokhin/Jeavons:2000,Bulatov/Krokhin/Jeavons:2001,Chen/Dalmau:laac,Dalmau:2000,Dalmau/Gavalda/Tesson/Therien:05,Dalmau/Pearson:1999,Feder:nearsubgroup,Feder/Vardi:1998,Jeavons/Cohen/Cooper:1998,Jeavons/Cohen/Gyssens:1997,Kirousis:1993},
a complete classification is still not known.

The other usual way to define subclasses of the general CSP is by
restricting the possible scopes, not the relations, that can appear
in a constraint. The state of affaires here is a way better: It has
been proved~\cite{Grohe:03} that, under certain plausible assumptions,
the tractable class identified in~\cite{Dalmau/Kolaitis/Vardi:02} is the only one, 
settling completely the question.

Our goal in this paper is to introduce a general condition, such that every constraint
language $\Gamma$ satisfying this condition, leads to a subclass of the CSP, $\csp(\Gamma)$,
solvable in polynomial time. In order to place our result in context it will be necessary to
include a short description of the cases of the CSP, known to be tractable.

In our classification we shall distinguish between {\em pure} and {\em hybrid} tractable cases.
Intuitively, {\em pure} tractable cases are those that are explained by a simple and concrete
combinatorial principle whereas {\em hybrid} tractable cases are those that can be reduced (not necessarily
without a considerable degree of sophistication) to a combination of pure cases. Let us recall here that this distinction is our rather personal attempt
to classify the tractable cases of the CSP and does not pretend to be any claim about the ``true nature'' of the tractable
cases.

According to our view there are basically three pure maximal
tractable cases: {\em width 1}, {\em bounded strict width}, and {\em
Mal'tsev} problems. This classification corresponds to the three
tractable families isolated in Feder and
Vardi~\cite{Feder/Vardi:1998}: width 1, bounded strict width, and
subgroup problems, in which the latter has been enlarged as to
include all Mal'tsev problems, proven to be tractable recently by
Bulatov~\cite{Bulatov:maltsev} (see also~\cite{Bulatov/Dalmau:05}).
The vast majority of tractable cases of the CSP identified in the
past (prior~\cite{Bulatov/Krokhin/Jeavons:2001}) fall in one of
these three pure categories. As an illustrative example consider the
six tractable cases in the boolean domain identified by
Schaefer~\cite{Schaefer:1978}, namely, $0$-valid, $1$-valid, Horn,
dual-Horn, bijunctive and affine problems. The first four classes
are particular instances of width 1 problems whereas the fifth and
sixth class belong to bounded strict width and Mal'tsev
respectively.

In~\cite{Bulatov/Krokhin/Jeavons:2001}, the pursuit of new tractable cases of the CSP took
a new direction. The new class identified in~\cite{Bulatov/Krokhin/Jeavons:2001}, the so-called paper-scissor-stone problems,
could be regarded as constituted by an amalgam of Horn problems (and hence width 1 problems) and
dual discriminator problems~\cite{Cooper/Cohen/Jeavons:1994}, known to be particular instances of bounded strict width problems.
The algorithm devised in~\cite{Bulatov/Krokhin/Jeavons:2001} exploits the fact that the interaction between  the width $1$ part and
the bounded strict width part of the instance is very constrained. The class of paper-scissor-stone problems
is the first hybrid tractable case.

All tractable cases identified
after~\cite{Bulatov/Krokhin/Jeavons:2001} with the notable exception
of Mal'tsev problems are hybrid problems (the
references~\cite{Bulatov:3element,Bulatov03:conservative,Bulatov04:graph,Bulatov:2semilattices,Chen/Dalmau:laac,Dalmau/Gavalda/Tesson/Therien:05}
constitute, up to the best of our knowledge, a complete list).
Indeed, it is not daring to say that is very likely that much of the
future progress in the study of the complexity of the CSP will come
from a better understanding of the interaction between the different
sources of pure tractability.

A unifying framework for tractability of constraint satisfaction has been developed by Jeavons and coauthors
in a sequence of papers culminating in~\cite{Jeavons/Cohen/Gyssens:1997}; the key theme of this framework is that it is generally
possible to explain the tractability of a certain subclass of the CSP, $\csp(\Gamma)$, by means of
certain algebraic invariance properties of the relations in $\Gamma$. Consider, for instance,
the class of bounded strict width problems: It is well known~\cite{Feder/Vardi:1998} (see also~\cite{Jeavons/Cohen/Cooper:1998}) that a
constraint language
$\Gamma$ is bounded strict width if an only if there exists a near-unanimity operation
$\varphi$, namely, an operation $\varphi:A^k\rightarrow A$ with $k\geq 3$ satisfying
\begin{center}
$\varphi(x,y,..,y)=\varphi(y,x,..,y)=\cdots=\varphi(y,y,..,x)=y$
\end{center}
for all $x,y\in A$, such that every relation in $\Gamma$ is invariant under $\varphi$.

 In a similar vein, a constraint
language $\Gamma$ is Mal'tsev if all its relations are invariant under a Mal'tsev operation, ie,
an operation $\varphi:A^3\rightarrow A$ satisfying
$$\varphi(x,y,y)=\varphi(y,y,x)=x, \text{ for all } x,y\in A.$$

Similar characterizations, in terms of algebraic invariance properties, are known for the vast majority of tractable
cases of the CSP. In fact, the connection between tractability of CSP and invariance properties is tighter,
as it can be shown that the complexity of a given subclass of the CSP, $\csp(\Gamma)$, depends {\em only}
on the the set of operations under which $\Gamma$ is invariant~\cite{Jeavons:1998}.

Generalized majority-minority
operations first arose in the study of the learnability of relatively quantified generalized formulas~\cite{Bulatov/Chen/Dalmau:04}.
An operation $\varphi:A^k\rightarrow A$ with $k\geq 3$ is a
\emph{generalized majority-minority (GMM) operation} if
for all $a,b \in A$,
\begin{center}
$\varphi(x,y,..,y)=\varphi(y,x,..,y)=\cdots=\varphi(y,y,..,x)=y$

\hspace{5cm} for all $x,y\in\{a,b\}$

or

$\varphi(x,y,..,y)=\varphi(y,y,..,x)=x \;\; \text{ for all } x,y\in\{a,b\}.$
\end{center}

GMM operations generalize both near-unanimity and Mal'tsev operations. Intuitively, an operation is GMM if for each $2$-element
subset of its domain acts either as a near-unanimity or as a Mal'tsev. 
In this paper we prove that every constraint language $\Gamma$ in which all its relations are
invariant under a GMM operation $\varphi$, gives rise to a subclass of the CSP solvable in polynomial-time.
This new family of constraint satisfaction problems includes among other all bounded strict width and
all Mal'tsev problems and, hence, constitutes one of the largest tractable classes of the CSP.
Our algorithm exploits a feature which is already used -at least implicitely- in the known algorithms for 
Malt'sev and Near-unanimity problems: for any relation $R$ invariant under a GMM operation $\varphi$ it is always possible
to obtain a ``succint'' representation, in the form of a relation $G\subseteq R$ that {\em generates} $R$, i.e., such that $R$ is the
smallest relation invariant under $\varphi$ containing $G$. Indeed, our algorithm for GMM problems can be viewed as
a mixture of the algorithms for Near-unanimity and Malt'sev problems. However, it should be point out that the interaction between
the two conditions is rather intricate.

\section{Preliminaires}

 Let
$A$ be a finite set and let 
$n$ be a positive integer.  A $n$-ary relation on $A$ is any subset of $A^n$.
In what follows, for every positive integer $n$, $[n]$ will denote the set $\{1,\dots,n\}$.

A constraint
satisfaction problem is a natural way to express simultaneous
requirements for values of variables. More precisely,

\begin{defi} An instance of a {\em constraint satisfaction problem} consists of:
\begin{itemize}
\item a finite set of variables, $V=\{v_1,\dots,v_n\}$;
\item a finite domain of values, $A$;
\item a finite set of constraints $\{C_1,\dots,C_m\}$;
each constraint $C_l,\; l\in[m]$, is a pair $((v_{i_1},\dots,v_{i_{k_l}}),S_l)$ where:
\begin{itemize}
\item $(v_{i_1},\dots,v_{i_{k_l}})$ is a tuple of variables of length $k_l$, called the {\em constraint scope} and
\item $S_l$ is an $k_l$-ary relation on $A$, called the {\em contraint relation}.
\end{itemize}
\end{itemize}
\end{defi}

A {\em solution} to a constraint satisfaction instance is
a mapping $s:V\rightarrow A$ such that for each constraint $C_l$,
$l\in [m]$, we have that $(s(v_{i_1}),\dots,s(v_{k_l}))$ 
belongs to $S_l$.
Deciding whether or not a given problem instance has a
solution is NP-complete in general, even when
the constraints are restricted to binary
constraints~\cite{Mackworth:1977} or the domain of the problem has
size 2~\cite{Cook:1971}. However by imposing restrictions on
the constraint relations it is possible to
obtain restricted versions of the problem that are tractable.

\begin{defi}
For any set of relations $\Gamma$, $\csp(\Gamma)$ is defined to be the
class of decision problems with:
\begin{itemize}
\item Instance: A constraint satisfaction problem instance ${\mathcal P}$,
  in which all constraint relations are elements of $\Gamma$.
\item Question: Does ${\mathcal P}$ have a solution?
\end{itemize}
\end{defi}

In the last few years much effort has been devoted to the
identification of those sets $\Gamma$ for which $\csp(\Gamma)$ is
solvable in polynomial time
(See~\cite{Bulatov:3element,Bulatov:maltsev,Bulatov03:conservative,Bulatov04:graph,Bulatov:2semilattices,Bulatov/Jeavons/Volkov:2001,Bulatov/Krokhin/Jeavons:2000,Bulatov/Krokhin/Jeavons:2001,Chen/Dalmau:laac,Dalmau:2000,Dalmau/Gavalda/Tesson/Therien:05,Dalmau/Pearson:1999,Feder:nearsubgroup,Feder/Vardi:1998,Jeavons/Cohen/Cooper:1998,Jeavons/Cohen/Gyssens:1997,Kirousis:1993}).
In order to isolate such ``islands of tractability'' it has been
particulary useful to consider certain closure conditions on the
relations of $\Gamma$. In order to make this more precise we need to
introduce the following definition, which constitues the cornerstone
of the so-called {\em algebraic approach} of the study of the CSP.

\begin{defi}
\label{def:preserves}
Let $\varphi:A^k\rightarrow A$ be an $k$-ary operation on $A$ and let $R$ be a $n$-ary
relation over $A$. We say that $R$ is invariant under $\varphi$ if
for all (not necessarily different) tuples ${\bf t_1}=(t^1_1,\dots,t^1_n),
\dots,{\bf t_k}=(t^k_1,\dots,t^k_n)$ in $R$, the tuple
$\varphi({\bf t_1},\dots,{\bf t_k})$ defined as
$$(\varphi(t^1_1,\dots,t^k_1),\dots,\varphi(t^1_n,\dots,t^k_n))$$
belongs to $R$.
\end{defi}

Given a relation $R$ and an operation $\varphi$, we denote by $\langle R\rangle_{\varphi}$
the smallest relation $S$ that contains $R$ and that it is invariant under $\varphi$.
Very often, the operation $\varphi$ will be clear from the context and we will
drop it writting $\langle R\rangle$ instead of $\langle R\rangle_{\varphi}$.

Let $\varphi:A^k\rightarrow A$ be any operation on $A$. We denote by $\inv(\varphi)$ the
set containg all relations on $A$ invariant under $\varphi$.

The vast majority of constraint languages $\Gamma$ such that $\csp(\Gamma)$ is in PTIME
can be expressed as $\inv(\varphi)$ for some operation
$\varphi$. We refer the reader to the references pointed out in the
introduction for a complete (up to the best of our knowledge) list
of operations that lead to a tractable class of the CSP. In what
follows we shall introduce only two families of operations, namely
near-unanimity and Malt'sev, which will be particularly relevant to
our work.

\begin{exa}{\bf (Near-Unanimity Operations)}
An operation $\varphi:A^k\rightarrow A$ with $k\geq 3$ is
near-unanimity (NU) if for all $x,y\in A$, we have that
$$\varphi(x,y,..,y)=\varphi(y,x,..,y)=\dots=\varphi(y,y,..,x)=y$$
Tractability of $\csp(\inv(\varphi))$ for any arbitrary near-unanimity
operation $\varphi$ was proved in~\cite{Feder/Vardi:1998} (See
also~\cite{Jeavons/Cohen/Cooper:1998}).
\end{exa}

Many well known tractable cases of the CSP, such as $2$-SAT, or
the family of CSP with implicative constraints~\cite{Kirousis:1993,Cooper/Cohen/Jeavons:1994} are, in fact, particular instances
of this general case.

\begin{exa}{\bf(Mal'tsev Operations)}
An operation $\varphi:A^3\rightarrow A$ is Mal'tsev if
for all $x,y\in A$, we have
$$\varphi(x,y,y)=\varphi(y,y,x)=x$$
In~\cite{Bulatov:maltsev} (see~\cite{Bulatov/Dalmau:05} for a simpler
proof), it was shown that for every Malt'sev operation $\varphi$,
$\csp(\inv(\varphi))$ is solvable in polynomial time.  This general
result encompasses some previously known tractable cases of the CSP,
such as CSP with constraints defined by a system of linear
equations~\cite{Jeavons/Cohen/Gyssens:1997} or CSP with near-subgroups
and its cosets~\cite{Feder/Vardi:1998,Feder:nearsubgroup}.
\end{exa}

The class of generalized majority-minority operations
generalizes both near-unanimity and Mal'tsev operations.

\begin{defi}
An operation $\varphi: A^k \rightarrow A$ with $k \geq 3$
is a \emph{generalized majority-minority (GMM) operation} if
for all $a, b \in A$, either
\begin{equation}\label{eqn1}
\begin{array}{r}
\varphi(x,y,..,y)=\varphi(y,x,..,y)=\cdots=\varphi(y,y,..,x)=y \\
\text{for all } x,y\in\{a,b\}
\end{array}
\end{equation}
or
\begin{equation}\label{eqn2}
\varphi(x,y,..,y)=\varphi(y,y,..,x)=x \;\;  \text{for all } x,y\in\{a,b\}
\end{equation}
\end{defi}

Generalized majority-minority operations were introduced in the study
of the learnability of relatively quantified generalized formulas~\cite{Bulatov/Chen/Dalmau:04}.

Let us fix a GMM operation on a set $A$. A pair $a,b\in A$ is said to
be a \emph{majority} pair if $\varphi$ on $a,b$ satisfies (\ref{eqn1}). It
is said to be a \emph{minority} pair if $\varphi$ satisfies
(\ref{eqn2}). If $a=b$ then we will say $\{a,b\}$ is
a majority pair.

In this paper we prove the following result:

\begin{thm}
\label{th1}
For every GMM operation $\varphi$, $\csp(\inv(\varphi))$ is solvable in polynomial time.
\end{thm}

The proof is given in Section~\ref{proof}.

\section{Signatures and Representations}

Let $A$ be a finite set, let $n$ be a positive integer, let ${\bf t}=(t_1,\dots,t_n)$ be a $n$-ary tuple of elements in $A$, and
let $i_1,\dots,i_j$ elements in $[n]$.  By $\pr_{i_1,\dots,i_j} {\bf t}$ we denote the tuple $(t_{i_1},\dots,t_{i_j})$.
Similarly, for every $n$-ary relation $R$ on $A$ and for every $i_1,\dots,i_j\in [n]$ we denote
by $\pr_{i_1,\dots,i_j} R$ the $j$-ary relation given by $\{\pr_{i_1,\dots,i_j} {\bf t} : {\bf t}\in R\}$.
Given a subset $I=\{i_1,\dots,i_j\}$ of $[n]$ with $i_1<i_2<\dots<i_j$ we shall use
$\pr_I R$ to denote $\pr_{i_1,\dots,i_j} R$.

Let $n$ be a positive integer, let $A$ be a finite set,
let ${\bf t}$, ${\bf t'}$ be $n$-ary tuples and let $(i,a,b)$ be any element in $[n]\times A^2$
with $a\neq b$.
We say that $({\bf t},{\bf t'})$
witnesses $(i,a,b)$ if
$\pr_{1,\dots,i-1} {\bf t}=\pr_{1,\dots,i-1} {\bf t'}$,
$\pr_i {\bf t}=a$, and $\pr_i {\bf t'}=b$.
We also  say that ${\bf t}$ and ${\bf t'}$ witness
$(i,a,b)$ meaning that $({\bf t},{\bf t'})$ witnesses $(i,a,b)$.

Let $\varphi:A^k\rightarrow A$, $k\geq 3$, be a GMM operation
and let $R$ be any $n$-ary relation on $A$ (not necessarily invariant under $\varphi$).
We define the {\em signature} of $R$ {\em relative to $\varphi$},
$\sig_R\subseteq [n]\times A^2$, as the set containing all those $(i,a,b)\in [n]\times A^2$,
with $\{a,b\}$ a minority pair witnessed by tuples in $R$, that is
$$\begin{array}{r}
\sig_R=\{ (i,a,b)\in [n]\times A^2:  \{a,b\} \text{ a minority pair }, \\
 \exists {\bf t},{\bf t'}\in R \text { such that }( {\bf t},{\bf t'}) \text{ witnesses } (i,a,b)\}\end{array}$$

Fix a non-negative integer $j$. A subset $R'$ of $R$ is called a {\em representation} of $R$ {\em relative to $\varphi$}
of order $j$ if $\sig_R=\sig_{R'}$ and for every $I\subseteq\{1,\dots,n\}$ with
$|I|\leq j$, $\pr_I R=\pr_I R'$. We also say that $R'$ is a $j$-representation of $R$ relative to $\varphi$.
Observe that for any
$n$-ary relation there exists a $j$-representation with size bounded above by
$2|\sig_R|+\sum_{I\subseteq [n], |I|\leq j} |\pr_I R|$. We call any
such representation, a {\em compact representation} of $R$.
Note: When the operation $\varphi$ is clear from the context we shall drop the ``relative to $\varphi$''
with the implicit understanding that signatures and representations are relative to $\varphi$.

\begin{exa}\label{E:three}
Let $A$ be a finite set, let $\varphi:A^k\rightarrow A$, $k\geq 3$, be a GMM operation and let $j$ and $n$ positive integers.
We shall construct a $j$-representation $R'$ of $R=A^n$.

Initially $R'$ is empty. Fix any arbitrary element $d$ in $A$.
First, observe that $\sig_{R}$ contains all $(i,a,b)$ in $[n]\times A^2$ where
$\{a,b\}$ is a minority pair. For each triple $(i,a,b)$ in $\sig_R$ we
add to $R'$ two tuples ${\bf t^i_a}$, ${\bf t^i_b}$, where ${\bf t^i_a}$ is the tuple that has
$a$ in its $i$th coordinate and $d$ elsewhere and, accordingly, ${\bf t^i_b}$ is the tuple that has $b$ in
its $i$th coordinate and $d$ elsewhere. Notice that $({\bf t^i_a},{\bf t^i_b})$ witnesses $(i,a,b)$.
Hence, after adding to $R'$ a corresponding pair ${\bf t^i_a},{\bf t^i_b}$ for each $(i,a,b)$ in $\sig_R$
we have that $\sig_R=\sig_{R'}$. In a second step we add for each $i_1,\dots,i_{j'}$ with $j'\leq j$
and $i_1<i_2<\dots<i_j$, and every $a_1,\dots,a_{j'}\in A$, the tuple ${\bf t^{i_1,\dots,i_{j'}}_{a_1,\dots,a_{j'}}}$
that has $a_l$ in its $i_l$th coordinate for each  $l\in[j']$, and $d$ elsewhere. It is easy to verify that we
obtain a relation $R'$ such that for all $I\subseteq[n]$ with $|I|\leq j$, $\pr_I R'=A^{|I|}=\pr_I R$.
Hence $R'$ is a representation of $R$ of order $j$. Observe that, indeed, $R'$ is a {\em compact
representation} of $R$.
\end{exa}

The algorithm we propose relies on the following lemma.

\begin{lem} Let $A$ be a finite set, let
$\varphi:A^k\rightarrow A$, $k\geq 3$ be a GMM operation, let $R$
be a relation on $A$ invariant under $\varphi$ and let $R'$
be a representation of $R$ of order $k-1$. Then $\langle R'\rangle = R$
\end{lem}

\proof Let $n$ be the arity of $R$. We shall show that for every
$i\in[n]$, $\pr_{1,\dots,i} \langle R'\rangle=\pr_{1,\dots,i} R$ by
induction on $i$.  The case $i\leq k-1$ follows from
$\pr_{1,\dots,k-1} R'=\pr_{1,\dots,k-1} R$.  So let $i\geq k$ and let
${\bf a}=(a_1,\dots,a_i)\in\pr_{1,\dots,i} R$.  By induction
hypothesis, for some $b_i$, the tuple ${\bf
a'}=(a_1,\dots,a_{i-1},b_i)$ belongs to $\pr_{1,\dots,i} \langle
R'\rangle$.  In what follows we shall denote $\pr_{1,\dots,i} \langle
R'\rangle$ as $S$.

We consider two cases.

\medskip
\noindent{\em Case 1.} $\{a_i,b_i\}$ is majority.
\smallskip

In this case we show that, for every $I\subseteq\{1,\ldots, i\}$,
$\pr_I {\bf a}\in\pr_I S$. We show it by induction on
the cardinality $m$ of $I$. The result is true for $m\leq k-1$
due to the fact that $R'$ is a $(k-1)$-representation of $R$.
It is also true for
every set $I$ that does not contain $i$, since ${\bf a'}\in S$
certifies it. Thus let $I=\{j_1,\dots,j_m\}$ be any set of indices
$1\leq j_1< j_2<\dots<j_m=i$ with $m\geq k$ and also let
${\bf a^*}=\pr_I {\bf a}$. To simplify the notation let
us denote ${\bf a^*}=(c_1,\dots,c_m)$. By
induction hypothesis, $\pr_I S$ contains the tuples
${\bf d_1}=(d_1,c_2,\dots,c_m)$, ${\bf d_2}=(c_1,d_2,c_3,\dots,c_m)$,
$\dots$, ${\bf d_m}=(c_1,\dots,c_{m-1},d_m)$ for some $d_1,\dots,d_m\in A$.
If for some $i$, $c_i=d_i$ then we are done. Otherwise,
we can assume that $d_m=b_i$ and $c_m=a_i$ and henceforth
$\{d_m,c_m\}$ is majority. If for some $j$, the pair $\{d_j,c_j\}$ is
minority then we are done, because
$\varphi({\bf d_j},{\bf d_j},\dots,{\bf d_j},{\bf d_m})={\bf a^*}$.
Otherwise, $\{d_j,c_j\}$ is majority for any $j$. In this case
we have $\varphi({\bf d_1},{\bf d_2}\dots,{\bf d_{k-1}},{\bf d_k})={\bf a^*}$.

\medskip
\noindent{\em Case 2.} $\{a_i,b_i\}$ is minority.
\smallskip

Since ${\bf a}$ and ${\bf a'}$ belong to $\pr_{1,\dots,i} R$ then
there exists some tuples ${\bf t},{\bf t'}\in R$ such that
$\pr_{1,\dots,i} {\bf t}={\bf a}$ and $\pr_{1,\dots,i} {\bf t'}={\bf
a'}$.  Consequently, ${\bf t}$ and ${\bf t'}$ witness $(i,a_i,b_i)$
and since $\sig_{R'}=\sig_R$ we can conclude that $\pr_{1,\dots,i} R'$
(and hence $S$) contains tuples ${\bf c}=(c_1,\dots,c_{i-1},a_i)$ and
${\bf c'}=(c_1,\dots,c_{i-1},b_i)$ witnessing $(i,a_i,b_i)$. We shall
show that ${\bf a}$ can be obtained from ${\bf a'}$, ${\bf c}$ and
${\bf c'}$ by applying operation $\varphi$.  We need first an
intermediate tuple; we define ${\bf d}=(d_1,\dots,d_i)$ as
$\varphi({\bf a'},{\bf c'},\dots,{\bf c'},{\bf c})$.  Finally we
obtain ${\bf e}=(e_1,\dots,e_i)$ as $\varphi({\bf a'},{\bf
a'},\dots,{\bf a'},{\bf d})$.  Let us see that the tuple obtained
${\bf e}$ is indeed ${\bf a}$. For each $l\in[i-1]$, if $\{a_l,c_l\}$
is a majority pair then $d_l=\varphi(a_l,c_l,\dots,c_l,c_l)=c_l$, and
consequently $e_l=\varphi(a_l,a_l,\dots,a_l,c_l)=a_l$. Otherwise, if
$\{a_l,c_l\}$ is a minority pair, then
$d_l=\varphi(a_l,c_l,\dots,c_l)=a_l$ and consequently
$e_l=\varphi(a_l,a_l\dots,a_l,a_l)=a_l$. Finally, let us look at the
value of $e_i$. Since $\{a_i,b_i\}$ is a minority pair we have that
$d_i=\varphi(b_i,b_i,\dots,b_i,a_i)=a_i$ and hence
$e_i=\varphi(b_i,b_i,\dots,b_i,a_i)=a_i$ and we are done.\qed

\section{Proof of Theorem~\ref{th1}}
\label{proof}

We prove Theorem~\ref{th1} by giving a polynomial-time algorithm that
 decides correctly whether a CSP($\inv(\varphi)$) instance has a
 solution.  The structure of the algorithm mimics that
 of~\cite{Bulatov/Dalmau:05}.

Let ${\mathcal P}=(\{v_1,\dots,v_n\} ,A,\{C_1,\dots,C_m\})$ be a
CSP($\inv(\varphi)$) instance which will be the input of the
algorithm.

For each $l\in\{0,\dots,m\}$ we define ${\mathcal P_l}$ as the CSP
instance that contains the first $l$ constraints of ${\mathcal P}$,
that is ${\mathcal
P}_l=(\{v_1,\dots,v_n\},A,\{C_1,\dots,C_l\})$. Furthermore, we shall
denote by $R_l$ the $n$-ary relation on $A$ defined as
$$R_l=\{(s(v_1),\dots,s(v_n)) : s \text{ is a solution of } {\mathcal P}_l)$$

In a nutshell, the algorithm introduced in this section computes for
each $l\in\{0,\dots,m\}$ a compact representation $R'_l$ of $R_l$. In
the initial case ($l=0$), ${\mathcal P_0}$ does not have any
constraint at all, and consequently, $R_0=A^n$. Hence, a compact
representation of $R_0$ can be easily obtained as in Example 3. Once a
compact representation $R'_0$ of $R_0$ has been obtained the algorithm
starts an iterative process in which a compact representation
$R'_{l+1}$ of $R_{l+1}$ is obtained from $R'_l$ and the constraint
$C_{l+1}$. This is achieved by means of a call to procedure {\tt
Next}, which constitutes the core of the algorithm.  The algorithm
then, goes as follows:

\begin{tabbing}
{\bf Algorithm} {\tt GMM}($(\{v_1,\dots,v_n),A,\{C_1,\dots,C_m\})$) \\
{\em Step 1} \ \ \ \=  {\bf set} $R'_0$ as in Example 3 \\
{\em Step 2} \> {\bf for each} $l\in\{0,\dots,m-1\}$ {\bf do} \\
 \> \ \ \ \ \= (let $C_{l+1}$ be $((v_{i_1},\dots,{v_{i_{k_{l+1}}}}),S_{l+1})$) \\
{\em Step 2.1}  \> \> {\bf set} $R'_{l+1}:=\text{\tt Next}(R'_l,i_1,\dots,i_{k_{l+1}},S_{l+1})$ \\
\> {\bf end for each} \\
{\em Step 3} \> {\bf if} $R'_m\neq \emptyset$ {\bf return yes} \\
{\em Step 4} \> \> {\bf otherwise return no}
\end{tabbing}

Observe that if we modify step 3 so that the algorithm returns an arbitrary tuple in $R'_m$ instead
of ``yes'' then we have an algorithm that does not merely solve the decision question but actually
provides a solution.

Correctness and polynomial time complexity of the algorithm are
direct consequences of the correctness and the running time of the
procedure {\tt Next}: As it is shown in Section~\ref{next}
(Lemma~\ref{le:next}) at each iteration of Step 2.1, the call $\text{\tt Next}(R'_l,i_1,\dots,i_{l+1},S_{l+1})$ 
correctly computes a compact
representation of the relation $\{ {\bf t}\in R_l :
\pr_{i_1,\dots,i_{l+1}} {\bf t}\in S_{l+1}\}$ which is indeed
$R_{l+1}$. Furthermore the cost of the call is polynomial in $n$,
$|A|$, and $|S_{l+1}|$, which gives as a total running time for the
algorithm polynomial (see Corollary~\ref{cor:solve} for a rough
approximation of the running time) in the size of the input. This
finishes the proof of the correctness and time complexity of the
algorithm, and hence, of Theorem~\ref{th1}.
\endproof

Let us remark that our emphasis here is on simplicity rather than in
efficiency. In fact, much better time bounds that the ones provided in
our analysis can be obtained by improving the code using data
structures and by performing a more accurate analysis of the running
time.

The remainder of the paper is devoted to defining and analyzing
procedure {\tt Next}.  In order to define procedure {\tt Next} it is
convenient to introduce previously a pair of procedures, namely {\tt
Nonempty} and {\tt Fix-values}, which will be intensively used by our
procedure {\tt Next}.

\subsection{Procedure {\tt Nonempty}}

This procedure receives as input a $k$-order compact representation
$R'$ of a relation $R$ invariant under $\varphi$, a sequence
$i_1,\dots,i_j$ of elements in $[n]$ where $n$ is the arity of $R$,
and a $j$-ary relation $S$ also invariant under $\varphi$. The output
of the procedure is either an $n$-ary tuple ${\bf t}\in R$ such that
$\pr_{i_1,\dots,i_j} {\bf t}\in S$ or ``no'' meaning that such a tuple
does not exist.

\begin{tabbing}
{\bf Procedure} {\tt Nonempty}($R',i_1,\dots,i_j,S$) \\
{\em Step 1} \ \ \ \= {\bf set} $U:=R'$ \\
{\em Step 2} \> {\bf while} $\exists {\bf t_1},{\bf t_2},\dots,{\bf t_k}\in U$ such that \\
\> \ \ \ \ \ \ \ \ \ \ $\pr_{i_1,\dots,i_j} \varphi({\bf t_1},{\bf t_2},\dots,{\bf t_k})\not\in \pr_{i_1,\dots,i_j} U$ {\bf do} \\
{\em Step 2.1} \> \ \ \ \= {\bf set} $U:=U\cup\{\varphi({\bf t_1},{\bf t_2},\dots,{\bf t_k})\}$ \\
\> {\bf endwhile} \\
{\em Step 3} \> {\bf if } $\exists {\bf t}$ in $U$ such that $\pr_{i_1,\dots,i_j} {\bf t}\in S$
{\bf then return} ${\bf t}$ \\
{\em Step 4} \> {\bf else return} ``no''
\end{tabbing}

We shall start by studying its correctness. First observe that every tuple in $U$ belongs initially to $R'$
(and hence to $R$), or it has been obtained by applying $\varphi$ to some tuples
${\bf t_1},{\bf t_2},\dots,{\bf t_k}$
that previously  belong to $U$. Therefore, since $R$ is invariant under $\varphi$, we can conclude that
during all the execution of the procedure $U\subseteq R$.
Consequently, if a tuple ${\bf t}$ is returned in step 3, then it belongs to $R$
and also satisfies that $\pr_{i_1,\dots,i_j} {\bf t}\in S$, as desired. It only
remains to show that if a ``no'' is returned in step 4 then there does not exist any tuple ${\bf t}$
in $R$ such that $\pr_{i_1,\dots,i_j} {\bf t}\in S$. In order to do this we need to
show some simple facts about $U$. Notice that at any point of the execution of the procedure $R'\subseteq U$. Then $U$ is also a representation
of $R$ and hence $\langle U\rangle=R$. Therefore we have that
$$\langle \pr_{i_1,\dots,i_j} U\rangle=\pr_{i_1,\dots,i_j} \langle U\rangle=\pr_{i_1,\dots,i_j} R$$
By the condition on the ``while'' of step 2 we have that when the
procedure leaves the execution of step $2$ it is necessarily the
case that for all ${\bf t_1},{\bf t_2},\dots,{\bf t_k}\in U$,
$\pr_{i_1,\dots,i_j} \varphi({\bf t_1},{\bf t_2},\dots,{\bf t_k})\in
\pr_{i_1,\dots,i_j} U$ and consequently $\pr_{i_1,\dots,i_j}
U=\langle \pr_{i_1,\dots,i_j} U\rangle=\pr_{i_1,\dots,i_j} R$.
Hence, if there exists some ${\bf t}$ in $R$ such that
$\pr_{i_1,\dots,i_j} {\bf t}\in S$ then it must exist some ${\bf
t'}$ in $U$ such that $\pr_{i_1,\dots,i_j} {\bf t'}\in S$ and we are
done.

Let us study now the running time of the procedure. It is only necessary to focus on steps 2 and 3.
At each iteration of the loop in step 2, cardinality of $U$ increases by one. So we can bound
the number of iterations by the size $|U|$ of $U$ at the end of the execution of the procedure.

The cost of each iteration is basically dominated by the amount of
computational time needed to check whether there exists some tuples
$\exists {\bf t_1},{\bf t_2},\dots,{\bf t_k}\in U$ such that
$\pr_{i_1,\dots,i_j} \varphi({\bf t_1},{\bf t_2},\dots,{\bf
t_k})\not\in \pr_{i_1,\dots,i_j} U$ in step 2. In order to try all
possible combinations for ${\bf t_1},{\bf t_2},\dots,{\bf t_k}$ in
$U$, $|U|^k$ steps suffice. Each one of these steps requires time
$O(|U| n)$, as tuples have arity $n$ and checking whether
$\varphi({\bf t_1},{\bf t_2},\dots,{\bf t_k})$ belongs to $U$ can be
done naively by a sequential search in $U$. Thus, the total running
time of step 2 is $O(|U|^{k+1} n)$. The cost of step $3$ is the cost
of finding a tuple ${\bf t}$ in $U$ satisfying $\pr_{i_1,\dots,i_j}
{\bf t}\in S$ which is $O(|U| |S| n)$. Putting all together we
obtain that the complete running time of the procedure is
$O(|U|^{k+2} n+|U| |S| n)$ which we can bound by $O(|U|^{k+2} |S|
n)$. So, it only remains to bound the size of $U$ (at the end of the
execution of the procedure). The size of $U$ can be bounded by the
initial size of $R'$ which is at most $O(n|A|^2+n^k |A|^k)=O((n
|A|)^k)$ (since $R'$ is compact) plus the number of iterations in
step 2, which is bounded by $|\pr_{i_1,\dots,i_j} R|$.

Consequently the total running time of the procedure can be bounded
by
$$O\left( \left((n|A|)^k +|\pr_{i_1,\dots,i_j} R|\right)^{k+2} |S| n\right
).$$ We want to remark here that the size of $\pr_{i_1,\dots,i_j} R$
can be exponentially large in the size of the input. For now we do
not deal with this issue. Later we shall see how, in order to
overcome this difficulty, we organize invoking to {\tt Nonempty} in
such a way that the value of $\pr_{i_1,\dots,i_j} R$ is conveniently
bounded.

\subsection{Procedure {\tt Fix-values}}

This procedure receives as input a compact representation $R'$ of a relation $R$
invariant under $\varphi$ and a sequence $a_1,\dots,a_m$, $m\leq n$ of elements of $A$ ($n$ is
the arity of $R$). The output is a compact representation of the relation given by
$$\{ {\bf t}\in R : \pr_1 {\bf t}=a_1,\dots,\pr_m {\bf t}=a_m\}$$

\begin{figure*}
\begin{tabbing}
{\bf Procedure} {\tt Fix-values}($R',a_1,\dots,a_m$) \\
{\em Step 1} \ \ \ \ \ \ \ \ \= {\bf set} $j:=0$; $U_j:=R'$ \\
{\em Step 2} \> {\bf while} $j<m$ {\bf do} \\
{\em Step 2.1} \> \ \ \ \= {\bf set} $U_{j+1}:=\emptyset$ \\
{\em Step 2.2} \> \> {\bf for each} $(i,a,b)\in[n]\times A^2$, $\{a,b\}$ is a minority pair, {\bf do} \\
{\em Step 2.2.1} \> \> \ \ \ \  {\bf if} \= $\text{\tt Nonempty}(U_j,j+1,i,\{(a_{j+1},a)\})\neq $''no'' and \\
\> \> \>       $(i,a,b)\in\sig U_j$ and $i> j+1$  {\bf then} \\
\> \> \> \ \ \ \ \= (let ${\bf t_1}$ be the tuple returned by $\text{\tt Nonempty}(U_j,j+1,i,\{(a_{j+1},a)\})$ \\
\> \> \> \>  and let ${\bf t_2}$,${\bf t_3}$ be tuples in $U_j$ witnessing $(i,a,b)$ )  \\
{\em Step 2.2.1.1} \> \> \> \> {\bf set} ${\bf t_4}:=\varphi({\bf t_1},{\bf t_2},\dots,{\bf t_2},{\bf t_3})$ \\
{\em Step 2.2.1.2} \> \> \> \> {\bf set} ${\bf t_5}:=\varphi({\bf t_1},{\bf t_1},\dots,{\bf t_1},{\bf t_4})$ \\
{\em Step 2.2.1.3}  \> \> \> \> {\bf set} $U_{j+1}:=U_{j+1}\cup\{{\bf t_1},{\bf t_5}\}$ \\
\> \> {\bf end for each} \\
{\em Step 2.3} \> \> {\bf for each } $k'\in [k-1]$  \\
\> \> \> {\bf for each} $l_1,\dots,l_{k'}\in [n]$ with $l_1<l_2<\dots<l_{k'}$  \\
\> \> \> \> {\bf for each} $d_1,\dots,d_{k'}\in A$ {\bf do} \\
{\em Step 2.3.1} \> \> \> \> \ \ \ \= {\bf if} $\text{\tt Nonempty}(U_j,l_1,\dots,l_{k'},j+1,\{(d_1,\dots,d_{k'},a_{j+1})\})\neq $''no'' {\bf then} \\
\> \> \> \> \> (let ${\bf t_6}$ be the tuple returned by the call \\
\> \> \> \> \> to $\text{\tt Nonempty}(U_j,l_1,\dots,l_{k'},j+1,\{(d_1,\dots,d_{k'},a_{j+1})\})$) \\
\> \> \> \> \> {\bf set} $U_{j+1}:=U_{j+1}\cup\{{\bf t_6}\}$ \\
\> \> {\bf end for each} \\
{\em Step 2.4} \> \> {\bf set} $j:=j+1$ \\
\> {\bf end while} \\
{\em Step 3} \> {\bf return $U_m$} \\
\end{tabbing}
\caption{Fix-values}
\label{Fix-values}
\end{figure*}

Figure~\ref{Fix-values} contains a description of the procedure.
Let us study its correctness. We shall show by induction on $j\in\{0,\dots,m\}$
that $U_j$ is a compact representation of $R_j=\{{\bf t}\in R : \pr_1 {\bf t}=a_1,\dots,\pr_j {\bf t} =a_j\}$.
The case $j=0$ is correctly settled in step 1. Hence it is only necessary to show that at every iteration
of the while loop in step 2, if $U_j$ is a compact representation of $R_j$ then $U_{j+1}$
is a compact representation of $R_{j+1}$. We shall start by showing that at the end
of the execution of step 2.2, $\sig_{U_{j+1}}=\sig_{R_{j+1}}$.
It is easy to see that
if any of the conditions of the ``if'' in step 2.2.1 is falsified then $(i,a,b)$ is not in $\sig_{R_{j+1}}$.
So it only remains to see that when the ``if'' in step 2.2.1 is satisfied, we have that (a)
$({\bf t_1},{\bf t_5})$ witnesses $(i,a,b)$, and
(b) ${\bf t_1}$ and
${\bf t_5}$ are tuples in $R_{j+1}$,

Proof of (a): We shall first show that for each $l\in[i-1]$, $\pr_l {\bf t_1}=\pr_l {\bf t_5}$.
Let  $c_l$ to be $\pr_l {\bf t_1}$ and let $d_l=\pr_l {\bf t_2}=\pr_l {\bf t_3}$. If $\{c_l,d_l\}$
is a majority pair then $\pr_l {\bf t_4}=\varphi(c_l,d_l,\dots,d_l,d_l)=d_l$  and hence
$\pr_l {\bf t_5}=\varphi(c_l,c_l,\dots,c_l,d_l)=c_l$. Otherwise,
if $\{c_l,d_l\}$ is a minority pair, then $\pr_l {\bf t_4}=\varphi(c_l,d_l,\dots,d_l,d_l)=c_l$ and
consequently $\pr_l {\bf t_5}=\varphi(c_l,c_l,\dots,c_l,c_l)=c_l$. So it only remains to show
that $\pr_i {\bf t_1}=a$ and $\pr_i {\bf t_5}=b$. We have $\pr_i {\bf t_1}=a$ as a direct consequence
of the fact that ${\bf t_1}$ is the tuple returned by the call to
 $\text{\tt Nonempty}(U_j,j+1,i,\{(a_{j+1},a)\})$. Observe also that as $({\bf t_2},{\bf t_3})$
witnesses $(i,a,b)$ we have that $\pr_i {\bf t_2}=a$ and $\pr_i {\bf t_3}=b$. Consequently,
since $\{a,b\}$ is a minority pair we have that
$\pr_i {\bf t_4}=\varphi(a,a,\dots,a,b)=b$ and hence $\pr_i {\bf t_5}=\varphi(a,a,\dots,a,b)=b$,
as desired.

Proof of (b): As ${\bf t_1}$ is the output of the call $\text{\tt Nonempty}(U_j,j+1,i,\{(a_{j+1},a)\})$, we can conclude that ${\bf t_1}$ belongs to $R_j$, $\pr_{j+1} {\bf t_1}=a_{j+1}$,
and $\pr_i {\bf t_1}=a$. Consequently ${\bf t_1}$ belongs to $R_{j+1}$. Furthermore, as ${\bf t_1}$, ${\bf t_2}$, and ${\bf t_3}$ are
in $R_j$ and $R_j$ is invariant under $\varphi$, we can conclude that ${\bf t_5}$ belongs to
$R_j$. Thus in order to see that ${\bf t_5}$ belongs to $R_{j+1}$ it only remains to
show that $\pr_{j+1} {\bf t_5}=a_{j+1}$. This can be obtained as a direct consequence of
(a), since as $({\bf t_1},{\bf t_5})$ witnesses $(i,a,b)$ and $i>j+1$, we have that
$a_{j+1}=\pr_{j+1} {\bf t_1}=\pr_{j+1} {\bf t_5}$.

We have just seen that at the end of step 2.2,
$\sig_{U_{j+1}}=\sig_{R_{j+1}}$. In Step 2.3, procedure {\tt
Fix-values} enlarges $U_{j+1}$ so that for every set $I\subseteq[n]$
with $|I|\leq k-1$, $\pr_I U_{j+1}=\pr_I R_{j+1}$. The proof of this
fact is rather straightforward. Let $k'$, $l_1,\dots,l_{k'}$,
$d_1,\dots,d_{k'}$ be the running parameters of a given interation
of step 2.3. It is easy to observe that the call to $\text{\tt
Nonempty}(U_j,l_1,\dots,l_{k'},j+1,\{(d_1,\dots,d_{k'},a_{j+1})\})$
is  different than ``no'' if and only if
$(d_1,\dots,d_{k'})\in\pr_{l_1,\dots,l_{k'}} R_{j+1}$. Furthermore,
if the call returns a tuple ${\bf t_6}$ then we can guarantee that
${\bf t_6}$ belongs to $R_{j+1}$ and that $\pr_I {\bf
t_6}=(d_1,\dots,d_{k'})$. We have just seen that at the end of the
execution of step 2.3 for every set $I\subseteq[n]$ with $|I|\leq
k-1$, $\pr_I U_{j+1}=\pr_I R_{j+1}$.

Notice that at each iteration of step 2.2, at most $2$ tuples are added for each $(i,a,b)$ in $\sig_{R_{j+1}}$.
Furthermore at each iteration of step 2.3, at most one tuple is added per each $k'$, $I$ with $|I|\leq k-1$
and tuple in $\pr_I R_{j+1}$. Consequently,
$U_{j+1}$ is compact. This completes the proof of its correctness.

Let us study now its time complexity. The ``while'' loop at step 2
is performed $m\leq n$ times. At each iteration the procedure
executes two loops (Step 2.2 and Step 2.3). The ``for each'' loop at
step 2.2 is executed for each $(i,a,b)$ in $[n]\times A^2$ with
$\{a,b\}$ a minority pair. That is a total number of times bounded
by $n |A|^2$. The cost of each iteration of the loop in Step 2.2 is
basically dominated by the cost of the call to procedure ${\tt
Nonempty}$ which costs $O(((n|A|)^k+|A|^2)^{k+2} n)$ which is $O((n
|A|)^{k(k+2)+1})$. The ``for each'' loop at step 2.3 is executed
$O((n |A|)^k)$ times. The cost of each iteration is basically the
cost of the call to {\tt Nonempty} which is
$O(((n|A|)^k+|A|^k)^{k+2}n)=O((n|A|)^{k(k+2)+1})$. Thus the total
cost of step 2.3 is $O((n|A|)^{k(k+2)+1+k})$. Thus the combined cost
of steps 2.2 and 2.3 is dominated by the cost of 2.3 which gives as
a total cost of $O((n|A|)^{k(k+2)+1+k})$ for each iteration of step
$2$. Since step $2$ is executed at most $n$ times we have a total
cost of the procedure of $O((n|A|)^{k(k+2)+1+k}n)$ which we shall
bound by $O((n|A|)^{(k+1)(k+2)})$.

\subsection{Procedure {\tt Next}}
\label{next}
We are now almost in a position to introduce procedure {\tt Next}.
Procedure {\tt Next} receives as input a compact representation $R'$ of a relation
$R$ invariant under $\varphi$, a sequence $i_1,\dots,i_j$ of elements in $[n]$
where $n$ is the arity of $R$, and a $j$-ary relation $S$ invariant under $\varphi$. The output of {\tt Next}
is a compact representation of the relation
$R^*=\{{\bf t}\in R : \pr_{i_1,\dots,i_j} {\bf t}\in S$\}.
It is an easy exercise to verify that $R^*$ must also be invariant under $\varphi$.

We shall start by defining a procedure, called {\tt Next-beta} that although equivalent
to {\tt Next} has a worse running time. In particular, the running time of {\tt Next-beta}
might be exponential to the arity $j$ of $S$ (recall that arities can be unbounded as 
we allow infinite constraint languages).

\begin{figure*}
\begin{tabbing}
{\bf Procedure} {\tt Next-beta}($R',i_1,\dots,i_j,S$) \\
{\em Step 1}  \ \ \ \= {\bf set} $U:=\emptyset$ \\
{\em Step 2} \> {\bf for each} $(i,a,b)\in[n]\times A^2$, $\{a,b\}$ is a minority pair {\bf do} \\
{\em Step 2.1} \> \ \  \= {\bf if } {\tt Nonempty}$(R',i_1,\dots,i_j,i,S\times\{a\})\neq$''no'' {\bf then} \\
\> \> \ \ \ \= (let ${\bf t_1}$ be {\tt Nonempty}$(R',i_1,\dots,i_j,i,S\times\{a\})$) \\
{\em Step 2.2} \> \> \> {\bf if} $\text{\tt Nonempty}(\text{\tt Fix-values}(R',\pr_1 {\bf t_1},\dots,\pr_{i-1} {\bf t_1}),i_1,\dots,i_j,i,S\times\{b\})\neq$''no'' \\
\> \> \> \ \ \= (let ${\bf t_2}$ be $\text{\tt Nonempty}(\text{\tt Fix-values}(R',\pr_1 {\bf t},\dots,\pr_{i-1} {\bf t}),i_1,\dots,i_j,i,S\times\{b\})$) \\
\> \> \> \> {\bf set} $U:=U\cup\{{\bf t_1},{\bf t_2}\}$ \\
\> {\bf end for each} \\
{\em Step 3} \> {\bf for each} $k'\in [k-1]$ \\
\> \> {\bf for each} $l_1,\dots,l_{k'}\in [n]$ with $l_1<l_2<\dots<l_{k'}$  \\
\> \> \> {\bf for each} $d_1,\dots,d_{k'}\in A$ {\bf do} \\
{\em Step 3.1} \> \> \> {\bf if } $\text{\tt Nonempty}(R',i_1,\dots,i_j,l_1,\dots,l_{k'},S\times\{(d_1,\dots,d_{k'})\})\neq$''no'' {\bf then} \\
\> \> \> \> (let ${\bf t_3}=\text{\tt Nonempty}(R',i_1,\dots,i_j,l_1,\dots,l_{k'},S\times\{(d_1,\dots,d_{k'})\})$) \\
\> \> \> \> {\bf set} $U:=U\cup\{{\bf t_3}\}$ \\
{\em Step 4} \> {\bf return} $U$ \\
\end{tabbing}
\caption{Next-beta}
\label{Next-beta}
\end{figure*}

Figure~\ref{Next-beta} contains a description of the procedure.
The overall structure of procedure {\tt Next-beta} is similar to that of procedure {\tt Fix-values}.
The procedure constructs a representation $U$ of $R^*$. Initially $U$ is empty. In step 2, {\text Next-beta} adds tuples
to $U$ so that when it leaves the execution of step 2, $\sig_U=\sig_R$. Let us analyze step 2. Observe that the condition of the ``if'' statement
$$\text{\tt Nonempty}(R',i_1,\dots,i_j,i,S\times\{a\})\neq\text{''no''}$$
of step 2.1 is satisfied if and only if there exists a tuple ${\bf t_1}\in R$ such that $\pr_{i_1,\dots i_j} {\bf t_1}\in S$ and
$\pr_i {\bf t_1}=a$. Hence if such a tuple does not exist then $(i,a,b)$ is not in $\sig_{R^*}$ and nothing needs to
be done for $(i,a,b)$. Now consider the condition of the ``if'' statement in step 2.2 which is given by
$$\begin{array}{c}\text{\tt Nonempty}(\text{\tt Fix-values}(R',\pr_1 {\bf t_1},\dots,\pr_{i-1} {\bf t_1}), \\
i_1,\dots,i_j,i,S\times\{b\})\neq\text{''no''}\end{array}$$

This condition is satisfied if and only if there exists some ${\bf t_2}$ in $R$ such that $\pr_{i_1,\dots,i_j} {\bf t_2}\in S$,
$\pr_{1,\dots,i-1} {\bf t_2}=\pr_{1,\dots,i-1} {\bf t_1}$ and $\pr_i {\bf t_2}=b$. It is immediate to see that if
the condition holds then ${\bf t_2}\in R^*$ and $({\bf t_1},{\bf t_2})$ witnesses $(i,a,b)$. It only remains to show that if $(i,a,b)\in\sig_{R^*}$ then
such a ${\bf t_2}$ must exist: Let
${\bf t_a},{\bf t_b}$ be tuples in $R^*$ witnessing $(i,a,b)$ and let ${\bf t_1}$ be
the tuple returned by the call to procedure {\tt Nonempty} in step 2.1. In order to prove the existence of
${\bf t_2}$ we shall use the usual trick. First define a tuple ${\bf u}$ as
$\varphi({\bf t_1},{\bf t_a},\dots,{\bf t_a},{\bf t_b})$ and finally let us define ${\bf t_2}$
as $\varphi({\bf t_1},{\bf t_1},\dots,{\bf t_1},{\bf u})$. Since ${\bf t_1},{\bf t_a},{\bf t_b}$
belong to $R^*$ and $R^*$ is invariant under $\varphi$ we can conclude that ${\bf t_2}$ belongs to $R^*$.
Let us show that $\pr_{1,\dots,i-1} {\bf t_2}=\pr_{1,\dots,i-1} {\bf t_1}$:
For each $l\in [i-1]$,  let $c_l$ be $\pr_l {\bf t_1}$ and let $d_l$ be $\pr_l {\bf t_a}=\pr_l {\bf t_b}$.
If $\{c_l,d_l\}$ is a majority pair then $\pr_l {\bf u}=\varphi(c_l,d_l,\dots,d_l,d_l)=d_l$ and
hence $\pr_l {\bf t_2}=\varphi(c_l,c_l,\dots,c_l,d_l)=c_l$. Otherwise, $\{c_l,d_l\}$ is a minority
pair and hence $\pr_l {\bf u}=\varphi(c_l,d_l,\dots,d_l,d_l)=c_l$.
Consequently, $\pr_l {\bf t_2}=\varphi(c_l,c_l,\dots,c_l,c_l)=c_l$ and we are done. Finally we
need to see that $\pr_i {\bf t_2}=b$. Observe that since $\{a,b\}$ is a minority pair
we have that $\pr_i {\bf u}=\varphi(a,a,\dots,a,b)=b$ and hence $\pr_i {\bf t_2}=\varphi(a,a,\dots,a,b)=b$.

We have just proved that, if $U$ is the representation output by
the procedure in Step 4, then $\sig_U=\sig_{R^*}$. It is
straightforward to verify that step 3 guarantees that for each $I$
with $|I|\leq k-1$, $\pr_I U=\pr_I R^*$. The analysis here is
basically identical to that of step 2.3 in procedure {\tt Fix-values}.
Consequently, $U$ is a representation of $R^*$.

At each iteration of step 2, at most $2$ tuples are added for each $(i,a,b)$ in $\sig_{R^*}$.
Furthermore at each iteration of step 3, at most one tuple is added per each $k'$, $I$, with $|I|\leq k-1$
and tuple in $\pr_I R_{j+1}$. Consequently,
$U$ is compact. This completes the proof of its correctness.

Let us study the running time of procedure {\tt Next-beta}. The loop of step $2$ is performed $n|A|^2$ times and the
cost of each iteration is basically the cost of steps 2.1 and 2.2 in which other procedures are called.
The cost of calling
{\tt Nonempty}$(R',i_1,\dots,i_j,i,S\times\{a\})\}$ in step 2.1 is $O(((n|A|)^k+r)^{k+2} |S| n)$
where $r$ is $|\pr_{i_1,\dots,i_j,i} R|$.

The cost of calling
$$\begin{array}{c}\text{\tt Nonempty}(\text{\tt Fix-values}(R',\pr_1 {\bf t_1},\dots,\pr_{i-1} {\bf t_1}), \\
i_1,\dots,i_j,i,S\times\{b\})\end{array}$$

in step 2.2 is the sum of the call to {\tt Fix-values} which is
$O((n|A|)^{(k+1)(k+2)})$ and the call to ${\tt Nonempty}$ which is
$O(((n|A|)^k+r)^{k+2} |S| n)$. Therefore,  the total cost of an
iteration of the loop of step $2$ is
$$O\left(\left((n|A|)^k+r\right)^{k+2} |S| \; n +
(n|A|)^{(k+1)(k+2)}\right)$$ Hence, in order to obtain the total
running time for the procedure we only need to multiply the previous
quantify by the number of iterations, which is $n|A|^2$, obtaining

$$n|A|^2\left(\left((n|A|)^k+r\right)^{k+2} |S| \; n +
(n|A|)^{(k+1)(k+2)}\right)$$

Let us take a closer look at the value of $r=|\pr_{i_1,\dots,i_j,i}
R|$. It is important to notice here that the set of possible
constraints $S$ that can appear in an instance is {\em infinite} and
henceforth it is not possible to bound the value of $j$.
Consequently, the value of $r$ might be exponential in the worst
case. However, it would be possible to bound the value of $j$ and
get a polynomial bound for $r$ if a {\em finite} subset $\Gamma$ of
$\inv(\varphi)$ is fixed beforehand and we assume that all
constraint instances use only constraint relations from $\Gamma$.
Such a situation is not completely unusual. In fact, a good number
of results on the complexity of $\csp(\Gamma)$ including the
pioneering work of Schaeffer~\cite{Schaefer:1978} assumes $\Gamma$
to be finite. By using the procedure {\tt Next-beta} it could be
possible to define a polynomial-time algorithm that solves
$\csp(\Gamma)$ for every finite subset $\Gamma$ of $\inv(\varphi)$.
However we are aiming here for a more general result. To this end,
we define a new procedure {\tt Next} which makes a sequence of calls
to {\tt Next-beta}.

\begin{tabbing}
{\bf Procedure} {\tt Next}($R',i_1,\dots,i_j,S$) \\
{\em Step 1} \ \= {\bf set} $l:=0$, $U_l:=R'$ \\
{\em Step 2} \> {\bf while} $l<j$ {\bf do} \\
{\em Step 2.1} \> \ \ \ \= {\bf set} $U_{l+1}:=\text{\tt Next-beta}(U_l,i_1,\dots,i_{l+1},\pr_{1,\dots,l+1} S)$ \\
{\em Step 2.2} \> \> {\bf set} $l:=l+1$ \\
\> {\bf end while} \\
{\em Step 3} \> {\bf return } $U_j$
\end{tabbing}

Observe that at each call of the procedure {\tt Next-beta} in step
2.1, the value of $r$ can be bounded by $|\pr_{1,\dots,l}
S||A|$, and hence the running time of each call to {\tt Next-beta}
can be bounded (very grossly) by
$$O\left(n^{(k+1)(k+2)+1}
|A|^{(k+1)(k+2)+2} |S|^{k+3}\right)$$

Finally the running time of {\tt Next} is obtained by multiplying by
$n$ (which always bounds $j$) the previous quantity.

\begin{lem}
\label{le:next} For every $n\geq 1$, every $n$-ary relation $R$
invariant under $\varphi$, every compact representation $R'$ of $R$,
every $i_1,\dots,i_j\in [n]$, and every $j$-ary relation $S$
invariant under $\varphi$, $\text{\tt Next}(R',i_1,\dots,i_j,S)$
computes a compact representation of $R^*=\{{\bf t}\in R :
\pr_{i_1,\dots,i_j}\in S\}$ in time $O\left((n|A|)^{(k+1)(k+2)+2}
|S|^{k+3}\right)$. Furthermore $R^*$ is invariant under $\varphi$.\qed
\end{lem}

Finally, we have

\begin{cor}
\label{cor:solve} Algorithm {\tt Solve} decides correctly if an
instance $\mathcal P$ of $\csp(\inv(\varphi))$ is satisfiable in
time $O(m(n|A|)^{(k+1)(k+2)+2} |S^*|^{k+3})$ where $n$ is the number
of variables of ${\mathcal P}$, $m$ is its number of constraints and
$S^*$ is the largest constraint relation occurring in ${\mathcal
P}$.\qed
\end{cor}

\section{Acknowledgments}
Research partially supported by the MCyT under grants TIC
   2002-04470-C03, TIC 2002-04019-C03, TIN 2004-04343, the EU PASCAL Network of Excellence,
   IST-2002-506778, and the
   MODNET Marie Curie Research Training Network, MRTN-CT-2004-512234.

\bibliographystyle{plain}
\bibliography{../../common/bibliography}

\end{document}

%% file: definitions.tex
\def\proof{\noindent{\bf Proof. }}

\def\marcafdem{\vrule width 1.2ex height 1.1ex depth 0.1ex}
\def\qed{\hbox{}\nobreak\hfill\hbox{\marcafdem}\par}

\def\endproof{\qed}

\long\def\BEGINCOMMENT #1\ENDCOMMENT{\relax}

\newcommand{\maps}\longrightarrow
\newcommand{\cmaps}\Longrightarrow

\newcommand{\csp}{\operatorname{CSP}}

\newcommand{\pr}{\operatorname{pr}}

%% file: defpart.tex
\newcommand{\sig}{\operatorname{Sig}}
\newcommand{\inv}{\operatorname{Inv}}